\documentstyle[epsfig,psfig]{mn}
\newif\ifAMStwofonts

\newcommand{\ha}{H$\alpha$}

\ifoldfss
\ifCUPmtlplainloaded \else
\NewTextAlphabet{textbfit} {cmbxti10} {}
\NewTextAlphabet{textbfss} {cmssbx10} {}
\NewMathAlphabet{mathbfit} {cmbxti10} {} 
\NewMathAlphabet{mathbfss} {cmssbx10} {} 
\fi
\ifAMStwofonts
\ifCUPmtlplainloaded \else
\NewSymbolFont{upmath} {eurm10}
\NewSymbolFont{AMSa} {msam10}
\NewMathSymbol{\upi} {0}{upmath}{19}
\NewMathSymbol{\umu} {0}{upmath}{16}
\NewMathSymbol{\upartial}{0}{upmath}{40}
\NewMathSymbol{\leqslant}{3}{AMSa}{36}
\NewMathSymbol{\geqslant}{3}{AMSa}{3E}

 \let\le=\leqslant
 \let\ge=\geqslant
\fi
\fi
\fi 

\ifnfssone
\addtoversion{normal}{\mathit}{cmr}{m}{it}
\addtoversion{bold}{\mathit}{cmr}{bx}{it}
\newmathalphabet{\mathbfit} 
\addtoversion{normal}{\mathbfit}{cmr}{bx}{it}
\addtoversion{bold}{\mathbfit}{cmr}{bx}{it}
\newmathalphabet{\mathbfss} 
\addtoversion{normal}{\mathbfss}{cmss}{bx}{n}
\addtoversion{bold}{\mathbfss}{cmss}{bx}{n}
\ifAMStwofonts
\ifCUPmtlplainloaded \else
\UseAMStwoboldmath
\makeatletter
\new@mathgroup\upmath@group
\define@mathgroup\mv@normal\upmath@group{eur}{m}{n}
\define@mathgroup\mv@bold\upmath@group{eur}{b}{n}
\edef\UPM{\hexnumber\upmath@group}
\new@mathgroup\amsa@group
\define@mathgroup\mv@normal\amsa@group{msa}{m}{n}
\define@mathgroup\mv@bold\amsa@group{msa}{m}{n}
\edef\AMSa{\hexnumber\amsa@group}
\makeatother
\mathchardef\upi="0\UPM19
\mathchardef\umu="0\UPM16
\mathchardef\upartial="0\UPM40
\mathchardef\leqslant="3\AMSa36
\mathchardef\geqslant="3\AMSa3E

 \let\le=\leqslant
 \let\ge=\geqslant
\fi
\fi
\fi 

\ifnfsstwo
\DeclareMathAlphabet{\mathbfit}{OT1}{cmr}{bx}{it}
\SetMathAlphabet\mathbfit{bold}{OT1}{cmr}{bx}{it}
\DeclareMathAlphabet{\mathbfss}{OT1}{cmss}{bx}{n}
\SetMathAlphabet\mathbfss{bold}{OT1}{cmss}{bx}{n}
\ifAMStwofonts
\ifCUPmtlplainloaded \else
\DeclareSymbolFont{UPM}{U}{eur}{m}{n}
\SetSymbolFont{UPM}{bold}{U}{eur}{b}{n}
\DeclareSymbolFont{AMSa}{U}{msa}{m}{n}
\DeclareMathSymbol{\upi}{0}{UPM}{"19}
\DeclareMathSymbol{\umu}{0}{UPM}{"16}
\DeclareMathSymbol{\upartial}{0}{UPM}{"40}
\DeclareMathSymbol{\leqslant}{3}{AMSa}{"36}
\DeclareMathSymbol{\geqslant}{3}{AMSa}{"3E}

 \let\le=\leqslant
 \let\ge=\geqslant
\fi
\fi
\fi 

\ifCUPmtlplainloaded \else
\ifAMStwofonts \else 
\def\upi{\pi}
\def\umu{\mu}
\def\upartial{\partial}
\fi
\fi

\title[Gas motions in NGC~3631]
{Gas motions in the plane of the spiral galaxy NGC~3631}

\author[A.M.Fridman et al.]
{A.M.~Fridman$^{1,\,2}$, O.V.~Khoruzhii$^{1,\,3}$,
E.V.~Polyachenko$^{1}$, A.V.~Zasov$^2$, \newauthor
O.K.~Sil'chenko$^2$,
A.V.~Moiseev$^2$, A.N.~Burlak$^{1,\,2}$, V.L.~Afanasiev$^4$,
\newauthor S.N.~Dodonov$^4$, J.H.~Knapen$^{5,6}$\\
$^1$Institute of Astronomy of the Russian Academy of Science, 48,
Pyatnitskaya St., Moscow, 109017, Russia\\
$^2$Sternberg Astronomical Institute, Moscow State University, University
prospect, 13, Moscow, 119899, Russia\\
$^3$National Research Center "Troitsk Institute for Innovation and
Thermonuclear Researches", \\
Troitsk, Moscow reg., 142092, Russia\\
$^4$Special Astrophysical Observatory of the Russian Academy of Sciences,
\\ Zelenchukskaya, 377140, Russia\\
$^5$Isaac Newton Group of Telescopes, Apartado 321, Santa Cruz
de La Palma, E-38700 Spain\\
$^6$Department of Physical Sciences,
University of Hertfordshire, Hatfield, Herts AL10 9AB}

\date{}
\pagerange{}
\pubyear{}

\begin{document}

\label{firstpage}

\maketitle

\begin{abstract}
{The velocity field of the nearly face-on galaxy NGC~3631, derived from
observations in the \ha\ line and H{\sc i} radio line, is analysed to
study perturbations related to the spiral structure of the galaxy. We
confirm our previous conclusion that the line-of-sight velocity field
gives evidence of the wave nature of the observed two-armed spiral
structure. Fourier analysis of the observed velocity field is used to
determine the location of corotation of the spiral structure of this
galaxy, and the radius of corotation $R_{\rm c}$ is found to be about
42$''$, or 3.2 kpc. The vector velocity field of the gas in the plane of
the disc is restored, and taking into account that we previously
investigated vertical motions, we now have a full 3D gaseous velocity
field of the galaxy. We show clear evidence of the existence of two
anticyclonic and four cyclonic vortices near corotation in a frame of
reference rotating with the spiral pattern. The centres of the
anticyclones lie between the observed spiral arms. The cyclones lie
close to the observed spirals, but their centres are shifted from the
maxima in brightness.}
\end{abstract}


\section{Introduction}

The branch of astronomy known as dynamics of galactic discs has
acquired, through long years of development, an unquestionably classical
status, but has so far evolved as a part of theoretical astrophysics
(e.g., Fridman \& Polyachenko 1984; Binney \& Tremaine
1987). Observational data, which are the basis for dynamical
investigations, have been up to now mostly one-dimensional: a mass
distribution in a disc is usually reconstructed from a surface
brightness profile and from a long-slit major-axis velocity profile or
rotation curve. Such an approach leads dynamicists to suppose a strict
axisymmetry of galactic discs. However, evidently the latter are not
axisymmetric in general. Particularly, bars and spiral arms are a clear
manifestation of such non-axisymmetry. Thus, in order to make dynamical
analyses more reliable, one needs two-dimensional (2D) data.

With the advent of CCD detectors, 2D photometric studies began to
appear. In particular, Kent (1984, 1985) has undertaken 2D decomposition
of CCD images of galaxies and has determined a lot of exponential disc
parameters. In a series of works Athanassoula and co-workers
(Consid\`ere \& Athanassoula 1988; Garc\'\i a G\'omez \& Athanassoula
1993) have used an azimuthal Fourier analysis of images of galactic discs
to reveal properties of their spiral structure, such as the number of
arms and their pitch angle. Two-dimensional velocity fields, however,
are rarely included into state-of-art dynamical investigations. The
maximum yield obtained from such data is usually a rotation curve
calculated in zero-order approximation of circular rotation, i.e. again
under the assumption of axisymmetry. However, 2D velocity fields contain
much more information.

The observed line-of-sight velocity of gas in spiral galaxies contains a
contribution not only of the regular rotation, but also of the velocity
perturbations due to the spiral density waves. In principle, an analysis
of the velocity field enables one to separate all these components of gas
motion, but this task is far from simple: the expected amplitude of the
main harmonics related to wave motion is about one order of magnitude
lower than the maximal velocity of rotation of a galaxy. In addition,
the presence of both nonplanar oscillations of the gas along the
rotation axis and local non-circular motions makes the observed velocity
field very complicated and difficult to interpret. Although
spiral-related perturbations of the gas motion were detected both in our
Galaxy (Yuan 1969 and references therein) and in many external galaxies
beginning with the classical work of Rots on M81 (1975), the amplitude
of perturbed velocities and pattern angular velocity are badly known
even for the best observed galaxies.

Different methods were proposed to determine kinematical parameters of
density waves from the observed line-of-sight velocity fields (see
Sakhibov \& Smirnov 1987, 1989, 1990; Bonnarel et al. 1988; Canzian,
1993; Sempere et al. 1995; Schoenmakers et al. 1997; Westpfahl 1998 and
references therein). However, all these methods have one or two
principal shortcomings. First, they are based upon an alleged
possibility to restore the equilibrium rotation velocity without the
analysis of the residual velocities. Such a possibility exists if we
deal with the results of a model experiment and know the form of the
gravitational potential (exactly the case with Canzian's 1993
investigation). However, when we analyse the line-of-sight velocity
field of a real galaxy, independent restoration of the rotation curve
becomes impossible (Lyakhovich et al. 1997; Fridman et al. 1997). A
rotation curve determined in the frame of a model of pure circular
motion has systematic errors of the order of the residual velocities,
and thus the residual field built on the basis of this curve does not
represent the real field of velocity perturbations caused by the density
wave (Lyakhovich et al. 1997; Fridman et al. 1997). Second, all
approaches mentioned above are based on the assumption of a 2D character
of the galactic motion in a disc, whereas any real galactic disc is a 3D
object and regular motions induced by a density wave are also three
dimensional in principle (Fridman et al. 1997; Fridman et al.,
1999). Thus the only direct approach to analyse the observed velocity
field is to seek for self--consistent solutions for the full vector
velocity field. In other words, the rotation velocity
and all three components of the perturbed velocity should be
determined simultaneously from the analysis of observational data,
taking into account the 3D nature of the galactic discs.

A recent attempt to restore the complete (three component) vector
velocity field in the gaseous discs of grand-design galaxies from the
observed field of line-of-sight velocities (Lyakhovich et al. 1997;
Fridman et al. 1997), gave us a hope to build, in the future, an
observationally-based foundation of the dynamics of the galactic discs.
The knowledge of the complete velocity field gives us at once (1)
rotation curve, (2) all the basic resonances: Lindblad and corotational,
and (3) knowledge of the residual velocity field, containing recently
discovered structures such as giant anticyclones (Fridman et al. 1997)
and cyclones (Fridman et al. 1999), the ``constituent parts" of the
spiral density waves. Finally, knowledge of the complete velocity field
helps to determine the collective process --- a kind of instability ---
which is responsible for the spiral--vortex structure of a given disc.
Thus, we can state without risk of exaggeration that the observed
velocity field provides the necessary observational base for the
construction of the dynamical portrait of a galaxy under consideration.

The aim of the present article is to restore and analyse the velocity
field of the gaseous disc of the grand--design galaxy NGC 3631, for
which two types of line-of-sight velocity data were obtained, well
complementing each other: in the radio H{\sc i} and optical H$\alpha$
lines (see below). The H{\sc i} observations used for this
study were obtained by Knapen (1997) with the Westerbork Synthesis Radio
Telescope, and the \ha\ observations were carried out at the Special
Astrophysical Observatory (SAO) with its 6-m reflector equipped with an
F/2.4 focal reducer and a scanning Fabry--Perot interferometer.

NGC~3631 is a rather bright non-barred galaxy with well-defined spiral
structure. Its optical axial ratio is close to unity: according to the
RC3 catalogue (Vaucouleurs et al. 1991), ${\rm log}~ a/b~=~0.02 \pm
0.07$, so this galaxy looks nearly face-on. Such an orientation is very
favourable for studying gas motions perpendicular to the plane of the
galaxy, which was the main topic of our previous paper (Fridman et al.,
1998, referred hereafter as Paper I).

In Paper~I we showed that non-circular gas motions in NGC~3631 have a
regular character, and that they are related to the observed two-armed
spiral structure, which has a wave nature. The vertical (that is
perpendicular to the plane of the disc) component of the gas motions as
revealed by a Fourier analysis method (Fridman et al. 1997), was also
found to be induced by the spiral density wave. The inclination angle of
the disc of NGC 3631 was found to be about $17^0$, which enables, using
the same observational data, the restoration of the vector velocity
field in the plane of this galaxy, which is the main objective of the
present paper.

NGC~3631 is a grand-design spiral galaxy of type SAc, at a distance of
15.4~Mpc, as estimated from its recession velocity using a Hubble
constant of 75~km\,s$^{-1}$\,Mpc$^{-1}$, which gives an angular scale of
75~pc per arcsec. Interestingly, the galaxy has been included in Arp's
(1966) atlas of peculiar galaxies, thanks to its ``straight arms'', and
``absorption tube crossing from inside to outside of southern
arm''. These features can be recognised in the $R$-image shown in
Fig.~1, but the galaxy as a whole looks to us rather normal. The atomic
hydrogen distribution has been described by Knapen (1997 and references
therein to earlier work), and the ionized hydrogen has been studied
through emission in the H$\alpha$ line by, among others, Boeshaar \&
Hodge (1977), Hodge (1982) and Rozas, Beckman \& Knapen (1996).

In Section~2 the results of a Fourier analysis of the observed
distributions of optical (\ha\ and $R$-band) surface brightness and of
H{\sc i} surface density are given, and compared with a Fourier analysis
of azimuthal distributions of the observed line-of-sight velocities.
Section~3 presents a model of large-scale gas motion, in which we assume
that the gas rotates in the galactic plane, and simultaneously
participates in the perturbed 3D--motions caused by the two-armed
density wave. The latter fact allows us to restrict the expansion of the
line-of-sight velocity field to the first three Fourier harmonics
($m_{\rm obs} = 1 - 3$) (Fridman et al. 1997). Two independent methods
are used to determine the positions of corotation and other resonances,
based on the relationships between the phases of azimuthal $F_\varphi$
and radial $F_r$ oscillations of the perturbed velocity and the phase
$F_\sigma$ of the perturbed surface density. In Section~4 we then
proceed to restore the vector velocity field of the gas in the plane of
the disc. We show that in a frame of reference corotating with the
spiral density wave pattern, giant cyclones exist alongside anticyclones
near the corotation radius. Anticyclones were described in our earlier
work (Fridman et al. 1997), whereas cyclones were predicted (Fridman
et al. 2000) in those gaseous galactic discs in which the gradient of
the azimuthal residual velocity exceeds the gradient of the rotation
velocity in the reference frame corotating with spirals. We briefly
summarize our main conclusions in Section~5.

We refer to Paper I for a description of the observations which are used
in this work.

\section{Spiral structure of NGC 3631}

\begin{figure*}
\hspace{-.2cm}\psfig{figure=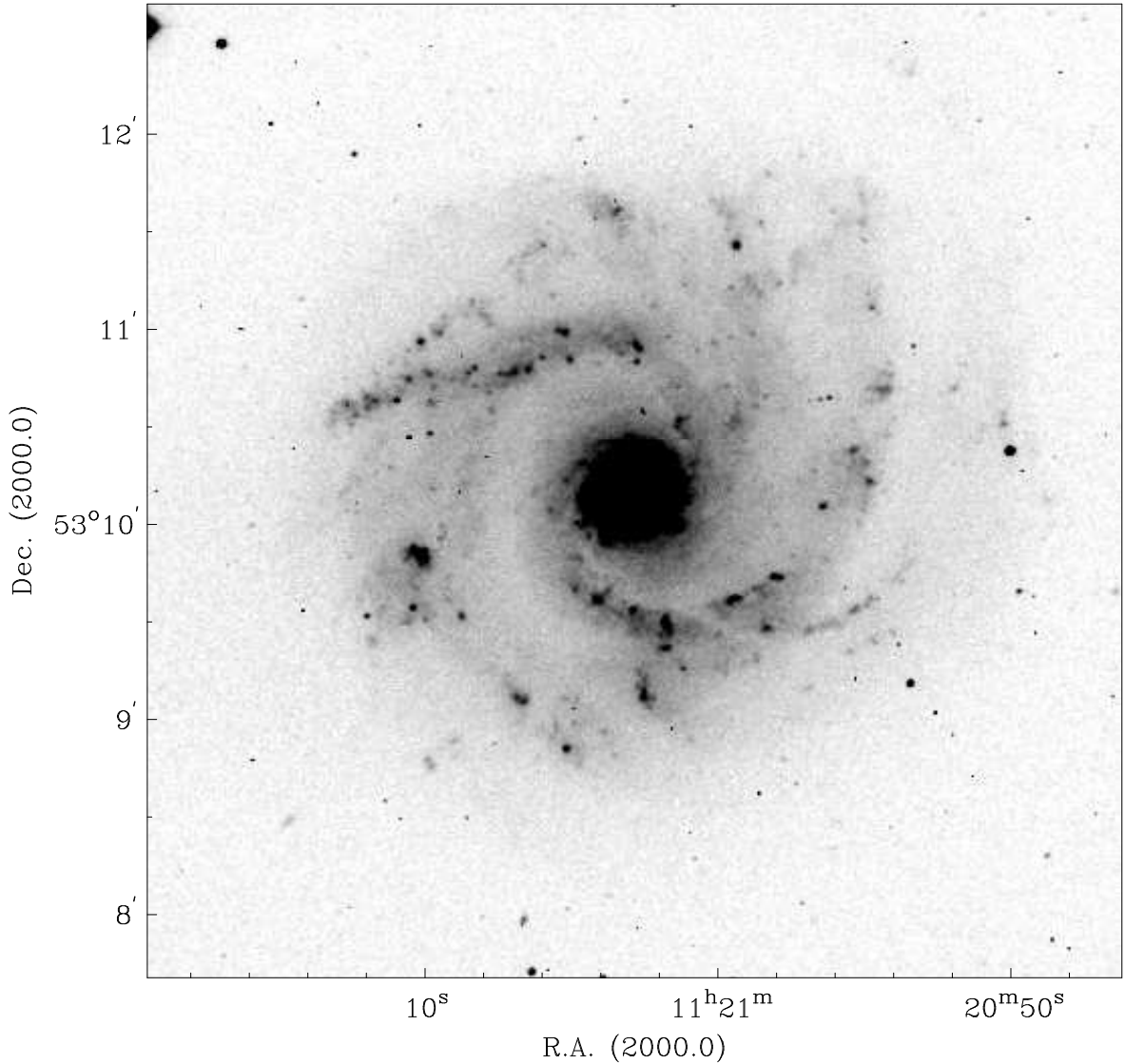,width=16cm}
\vspace{-.2cm}
\caption{$R$-band image of NGC~3631 as obtained from the ING archive}
\end{figure*}

\begin{figure}

\psfig{figure=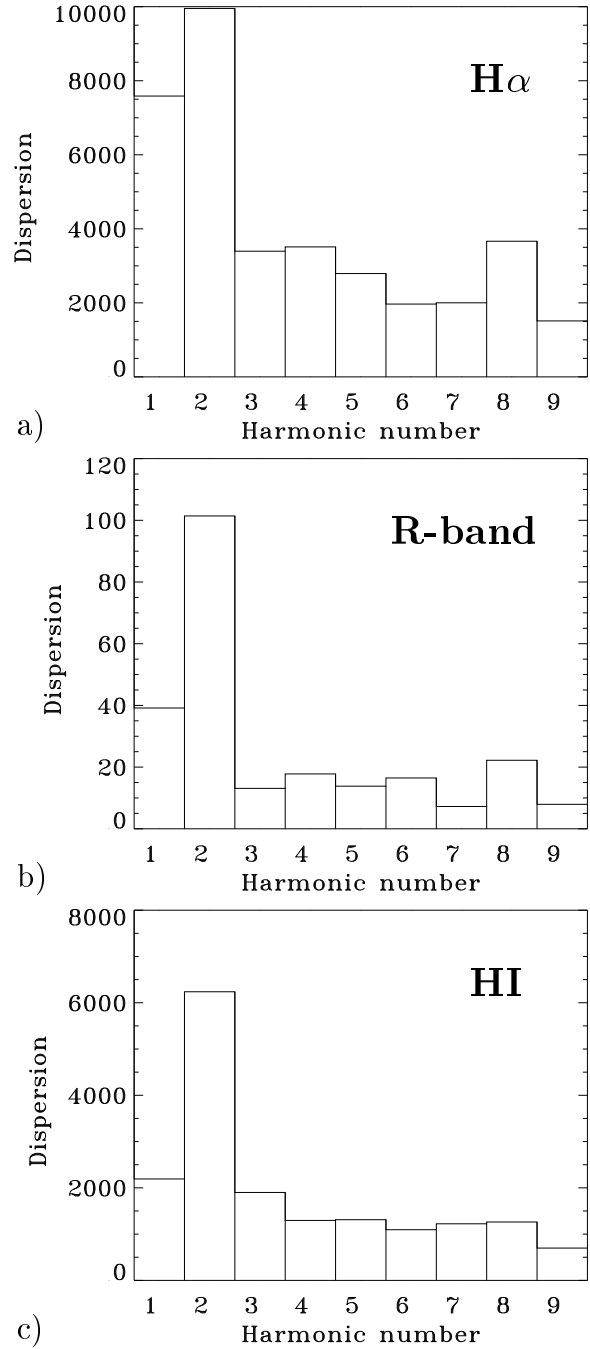,width=8cm,
bbllx=105pt, bblly=195pt, bburx=320pt, bbury=680pt, clip=}

\caption{Contribution of individual Fourier harmonics to the deviation,
or dispersion, of the brightness distribution from axial symmetry, as
derived from (a) our \ha\ image (Paper I), (b) our $R$-band image (ING
archive), and (c) 21 cm map (Knapen 1997).}
\label{f-1}
\end{figure}

Although the two-armed spiral structure of NGC 3631 is well defined in
optical light, it has a number of irregularities, especially in the
outer parts of the disc. The H{\sc i} map has much lower angular
resolution than optical images, yet it also clearly shows evidence of
the two-armed spiral structure (Knapen 1997). For the purpose of our
study we should be convinced that the second harmonic of the brightness,
$m_b~=~2$, of the spiral structure exceeds all other harmonics. To check
this, we divided the galactic disc into elliptical rings, corresponding
to circular rings after deprojection, and carried out a Fourier analysis
of the azimuthal brightness distribution, using an \ha\ image of NGC
3631 obtained at the SAO 6-m telescope through interferometric
observations (Paper~I), an $R$-band image of the galaxy, as obtained
from the Isaac Newton Group (ING) archive, and the neutral hydrogen
distribution, as obtained from 21~cm observations (Knapen 1997). We show
the $R$-band image, taken with the 1-m Jacobus Kapteyn Telescope, in
Fig.~1, which outlines the main spiral arm structure in this
galaxy. Technical information on this image can be found in Knapen
(1997). The series of histograms in Fig.~\ref{f-1} shows the
contributions of the different Fourier harmonics to the deviation from
an axially symmetrical distribution of brightness, or dispersion, for
the three images mentioned above. Throughout the paper, we restrict our
consideration to the region of the galactic disc $R$ $<$ $80''$ which
corresponds to the extent of the optical spirals.

\begin{figure}

\psfig{figure=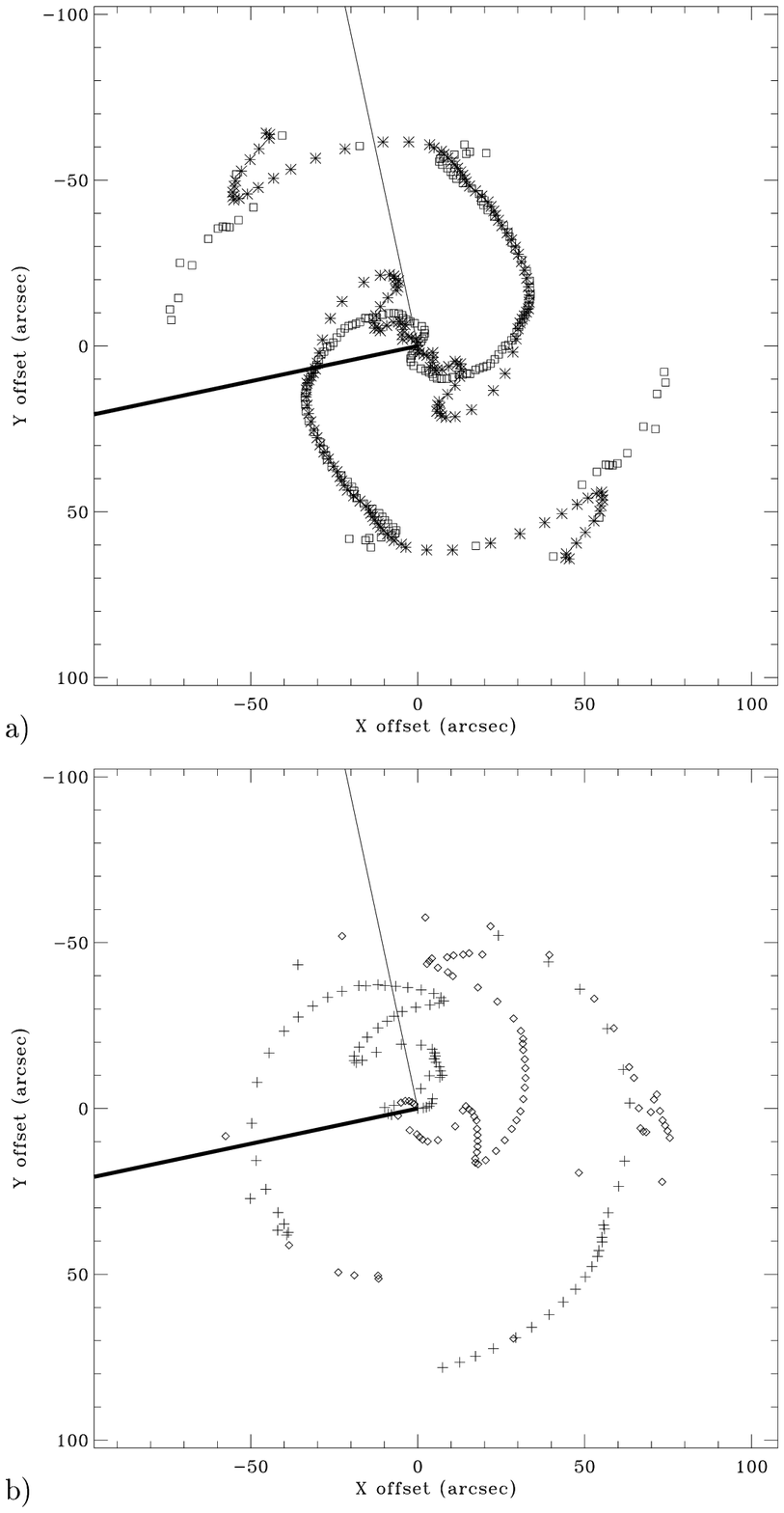,width=8cm,
bbllx=105pt, bblly=95pt, bburx=430pt, bbury=710pt, clip=}

\caption{Superposition of the positions of the maximum value of the
harmonics in the $R$ (asterisks) and \ha\ (squares) images: (a) top
panel, second harmonics; (b) lower panel, first harmonics.}
\label{f-2}
\end{figure}

Fig.~\ref{f-1} shows clearly that the second harmonic, which corresponds
to the observed two-armed structure, indeed dominates the spectrum. The
high level of the first harmonic in the \ha\ image is caused by the
non-symmetrical distribution of star-forming regions in the spiral arms,
and does not reflect the true contribution of the first Fourier harmonic
to the mass distribution in the galaxy. This assumption is supported by
the low level of the first harmonic in the $R$-band image and the H{\sc
i} map. To demonstrate this yet more clearly, we show in Fig.~\ref{f-2}
the existence of a tight correlation between lines of maximum values of
the second harmonic in the $R$- and \ha\ images (top panel), and a very
weak correlation between points and lines of maximum values of the first
harmonics (bottom panel).

\begin{figure}

\psfig{figure=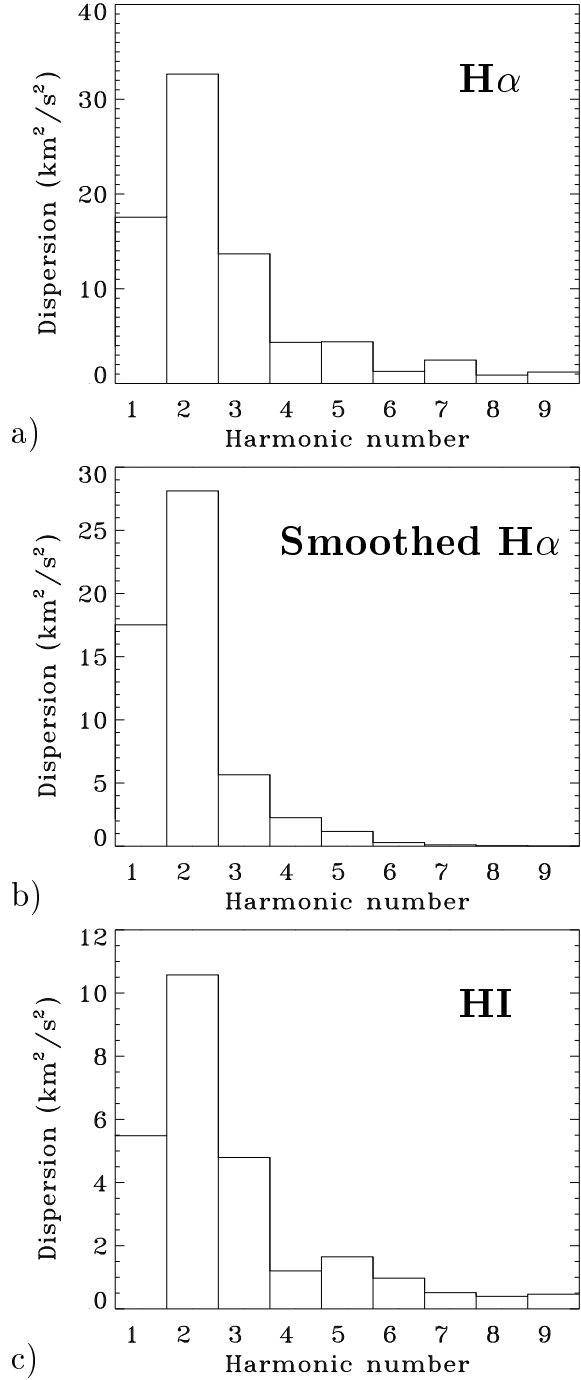,width=8cm,
bbllx=105pt, bblly=195pt, bburx=315pt, bbury=680pt, clip=}

\caption{Contribution of individual Fourier harmonics of the
line-of-sight velocity field into dispersion in the model of pure
circular motion in the region $R$ $<$ $40''$, as derived from (a)
original \ha\ line-of-sight velocity field, (b) \ha\ velocity field
smoothed to a resolution comparable to the 21 cm data, (c) 21 cm
velocity field.}
\label{f-3}
\end{figure}

\begin{figure}

\psfig{figure=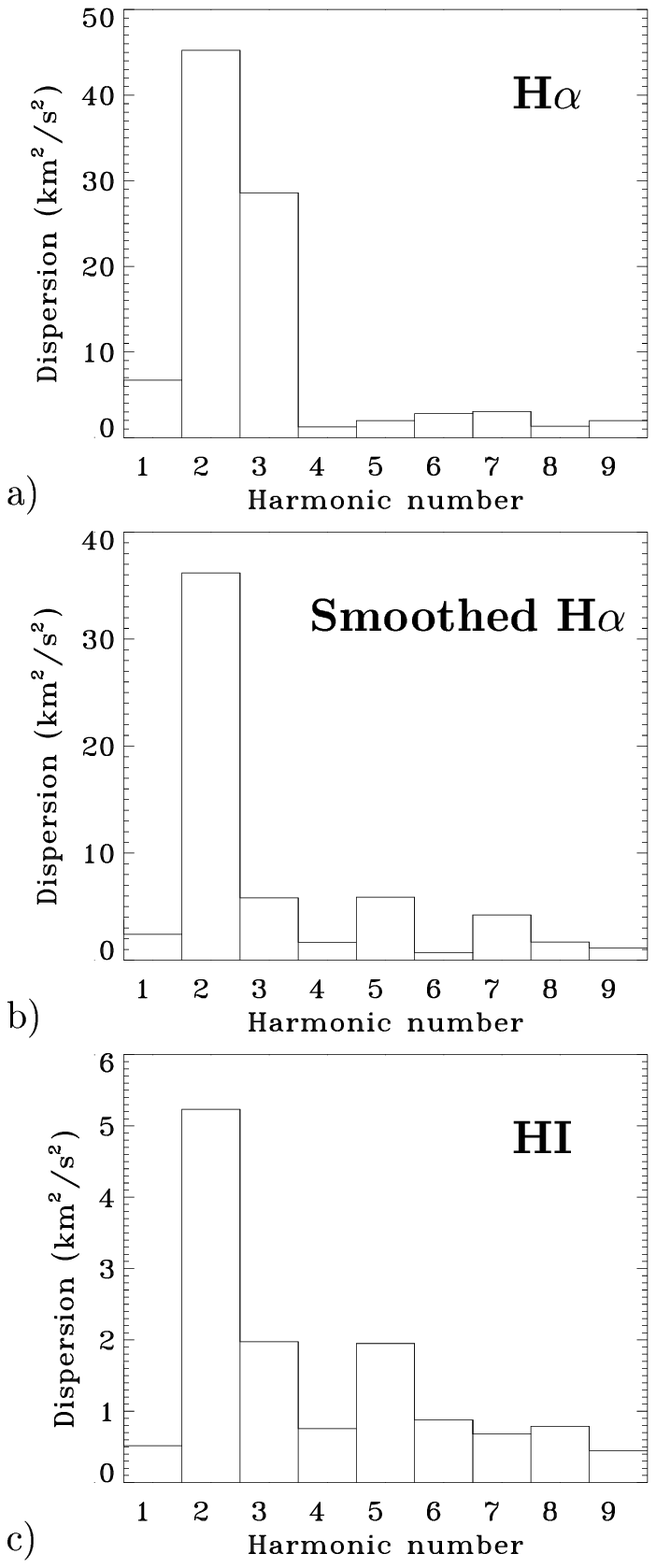,width=8cm,
bbllx=105pt, bblly=195pt, bburx=315pt, bbury=680pt, clip=}

\caption{As Fig.~4, now for the region $40''$ $<$ $R$ $<$ $80''$}
\label{f-3aa}
\end{figure}

As shown earlier (Sakhibov \& Smirnov 1989; Canzian 1993; Fridman
et al. 1997), if the circular velocity of gas in a galaxy is perturbed
by a two-armed spiral pattern, this must lead to the appearance of the
first and third Fourier harmonics ($m_{\rm obs}$ $=$ 1 and 3) in the
azimuthal distribution of the observed line-of-sight velocity. In
addition, the second harmonic ($m_{\rm obs}$ $=$ 2) may also appear if
the density wave induces vertical oscillations of the gas (Fridman
et al. 1997; Fridman et al. 1998). The predominance of the first three
harmonics in the line-of-sight velocity field of the galaxy is clear, as
first demonstrated in paper I. Here, we show this result for both optical
and radio velocity measurements in Fig.~\ref{f-3} and \ref{f-3aa}, using
$PA$ $=$ $336^\circ$, which gives minimum dispersion in a model of
pure circular motion for the radio line-of-sight velocity data in the part
of the galaxy under consideration. The optical data are practically
insensitive to the change of $PA$ by less than $10^\circ$. Fig.~\ref{f-3}
shows the contributions of different Fourier harmonics to the dispersion
in the model of pure circular motion, averaged over the inner part of the
disc ($R$ $<$ $40''$), whereas Fig.~\ref{f-3aa} shows similar histograms
for the outer part of the optical disc ($40''$ $<$ $R$ $<$ $80''$).

Comparison of the \ha\ (Figs.~\ref{f-3}\,a and \ref{f-3aa}\,a) with the
radio data (Figs.~\ref{f-3}\,c and \ref{f-3aa}\,c) shows that the
amplitude of the harmonics is much smaller in the latter than in the
former case. Such a difference could be a reason to distrust the
optical or radio data, if these data were related to the same velocity
field, but this is, in fact, not the case.

As pointed out in Paper I, reasons for differences between radio and
optical estimates of the amplitudes of the second Fourier harmonics on
the one hand, and of the first and the third harmonics on the other
hand, should be different. In the latter case, the difference may be
caused by the low resolution of the radio data. This is well
illustrated in Figs. 4b and 5b, where the amplitudes were calculated
after smoothing of the optical velocity field to a resolution of 14$''$,
close to that of the radio data. The squares of the amplitudes of the third
harmonic in Fig. 4b and of the first and third harmonics in Fig. 5b are
one-third/one-fifth as many as those in the original \ha\  data.
It thus follows from Figs. 4 and 5 that the histograms of the smoothed
\ha\ data occupy an intermediate position between histograms of
original \ha\  and H{\sc i} data --- amplitudes in the histograms b)
is closer to those in the histograms c) than those in histograms a). The
second harmonic of the smoothed \ha\   line-of-sight velocity field is
naturally much higher than the second harmonic of the H{\sc i} data. This
can be explained by the different optical depth of the gaseous disc in the
\ha\  and 21 cm lines, if one takes into account that the second harmonic
is caused by the vertical motions in the density wave which are
antisymmetrical with respect to the central plane of the disc (see
discussion in Paper~I).

It follows from Figs.~\ref{f-3} and \ref{f-3aa} that the contribution of
harmonics with $m_{\rm obs}$ $>$ 3 into the observed line-of-sight
velocity field is not significant for this galaxy. Note that this
conclusion is not a universal rule - our preliminary analysis of data
for other objects shows that it might be wrong for some other galaxies,
even though they have a two-armed grand-design spiral pattern. The
anomalously large value of the amplitude of the 5th harmonic in
Fig.~\ref{f-3aa}\,c is perhaps caused by the coincidence of the typical
spatial scales of the 5th harmonic in the outer part of the disc with
the resolution of the radio data. This hypothesis is supported by the
appearance of an artificial anomalous amplitude of the fifth harmonic in
the smoothed \ha\  data (Fig.~\ref{f-3aa}\,b).

\begin{figure}

\psfig{figure=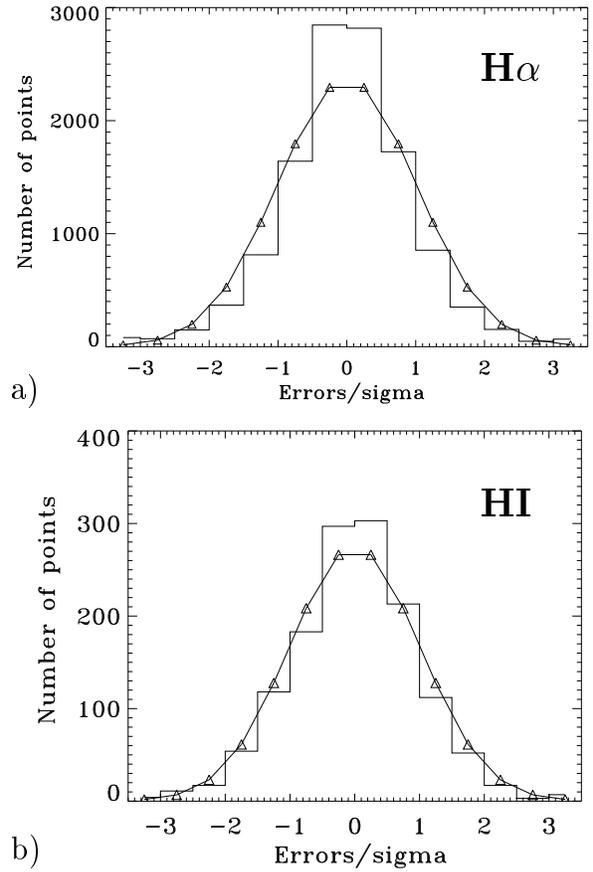,width=8cm,
bbllx=105pt, bblly=280pt, bburx=320pt, bbury=595pt, clip=}

\caption{Histogram of residuals (differences between the observed and
the model line-of-sight velocity fields) in the model of the
line-of-sight velocity field which takes into account only the first
three Fourier harmonics. The line shows the $\chi$-distribution with a
dispersion equal to that of the model. (a) For the original,
unsmoothed, \ha\ line-of-sight velocity field, (b) for the 21 cm
velocity field.}
\label{f-4}
\end{figure}

The extraction of the first three harmonics from the line-of-sight
velocity field reveals that the residuals have a non-regular noisy-like
character. Fig.~\ref{f-4} shows a histogram of the distribution of the
velocity residuals with $m$ $>$ $3$ all over the disc within the
deprojected radius $R$ $<$ $80''$. It can be satisfactorily
approximated by a Gaussian with a dispersion $\sqrt{\sigma^2}$ of about
10 km/s, which is close to the mean pixel-to-pixel error in the
individual velocity measurements. This allows the construction of a
simplified model of the velocity field of the galaxy, which uses only
the first and the third harmonics to describe the gas motion in the
plane of the disc.

All results given below are restricted to this model. The second
harmonic of the line-of-sight velocity field will be ignored here
because it relates to the vertical gas motions and was considered
earlier (see Paper I for details).

\section{Phase relationships and the position of corotation}

The method used to restore the vector velocity field and to determine
the corotation radius $R_{\rm c}$ from observations of line-of-sight
velocity distributions was described in detail by Lyakhovich et al.
(1997) and applied to some galaxies by Fridman et al. (1997) and Fridman
et al. (2000). The method is based on the comparison
of Fourier coefficients of azimuthal distributions of the observed
line-of-sight velocity, with those expected for a model where the
perturbed velocity components at the particular moment of time $\tilde
V_r$, $\tilde V_\varphi$, $\tilde V_z$ are caused by a two-armed spiral
density wave:
\begin{equation}
\label{eq:sprvz1}
\tilde{V_r}(R,\,\varphi ) = C_r(R) \, \cos [ 2\varphi- F_r (R) ] \, ,
\end{equation}
\begin{equation}
\label{eq:sprvz2}
\tilde{V_\varphi}(R,\,\varphi ) = C_\varphi(R) \, \cos [ 2\varphi-
F_\varphi (R) ] \, ,{\rm and}
\end{equation}
\begin{equation}
\label{eq:sprvz3}
\tilde{V_z}(R,\,\varphi ) = C_z(R) \, \cos [2\varphi - F_z (R)] \, ,
\end{equation}
where $C_i(R)$ and $F_i (R)$ are an amplitude and phase of i-component of
velocity.

The main idea behind this is the following. The
line-of-sight velocity is connected with the velocity components of the
gas by the relationship (Lyakhovich et al. 1997; Fridman et al. 1997)
\[
V^{obs} (R, \varphi )~=~ V_s ~+~ V_\varphi(R,\varphi ) \, \cos \varphi
\, \sin i ~+~
\]
\begin{equation}
\label{eq:vobsa}
~+~ V_r (R,\varphi) \, \sin \varphi \, \sin i ~+~ V_z
(R,\varphi) \,\cos i \, ,
\end{equation}
where $V_s$ is the systemic velocity of the galaxy.
Taking into account that $V_r$ $=$ $\tilde V_r$, $V_\varphi$ $=$
$V_{rot}$ $+$ $\tilde V_\varphi $, $V_z$ $=$ $\tilde V_z$,
where $V_{rot}$ is the rotation velocity, and substituting
(\ref{eq:sprvz1}) -- (\ref{eq:sprvz3}) in (\ref{eq:vobsa}) we obtain the
model representation of the line-of-sight velocity:
\[
V^{mod}(R,\,\varphi ) \,=\, V_s \,+\, \sin i \,[a_{1}(R) \, \cos \varphi
\,+\, b_{1}(R) \, \sin \varphi\,+\, \]

\[
~ +\, a_2(R) \, \cos 2\varphi \,+\, b_2(R) \, \sin 2 \varphi \,+\,
\]
\begin{equation}
\label{eq:vobsout20}
~+\, a_3 (R) \, \cos 3\varphi \,+\, b_3 (R) \, \sin 3 \varphi ]
\, ,
\end{equation}
where Fourier coefficients related to phases and
amplitudes of the velocity components are:
\begin{equation}
\label{eq:aAbB1}
a_1 ~=~ V_{rot}~+~\frac {C_r \sin F_r + C_\varphi \cos
F_\varphi } 2 \, ,
\end{equation}
\begin{equation}
\label{eq:aAbB2}
b_1 ~=~ -~ \frac{C_r \cos F_r - C_\varphi \sin F_\varphi }2 \, ,
\end{equation}
\begin{equation}
\label{eq:aAbB3}
a_2 ~=~ C_z \cos F_z \,{\rm cot}\, i \, ,
\end{equation}
\begin{equation}
\label{eq:aAbB4}
b_2 ~=~ C_z \sin F_z \,{\rm cot}\, i \, ,
\end{equation}
\begin{equation}
\label{eq:aAbB5}
a_3 ~=~ -~ \frac{C_r \sin F_r - C_\varphi \cos F_\varphi} 2 \, ,
\end{equation}
\begin{equation}
\label{eq:aAbB6}
b_3~ =~ \frac{C_r \cos F_r + C_\varphi \sin F_\varphi }2 \, .
\end{equation}

Calculating the Fourier coefficients of the observed line-of-sight
velocity field ($a_i^{obs}$, $b_i^{obs}$, with $i=$ 1, 2, 3) and
equating them to the model ones (Eqs. \ref{eq:aAbB1}-\ref{eq:aAbB6}), we
obtain the base to determine the amplitudes and phases of all three
velocity components.

The Fourier coefficients $a_1^{obs}$, $b_1^{obs}$, $a_2^{obs}$,
$b_2^{obs}$, $a_3^{obs}$ and $b_3^{obs}$ as well as the best fit
parameters of the galactic disc ($PA$, inclination, and centre position)
may be found from the observed line-of-sight velocity distribution
$V_j^{obs} (r_j,\varphi_j)$ by minimising in each elliptical ring
the quantity $\chi^2(R_k)$, as determined by the following
equation (see Lyakhovich et al. 1997; Fridman
et al. 1997 for details):
\[
\chi^2(R_k)~=~\sum_{j} \,\left(V_j (R_j,\varphi_j) ~-~ V_{s} ~-~ \right.
\]
\begin{equation}
\label{eq:leastsqr1}
\left.~~~-~ \sum_{n=1}^{n=3}\, \left[ a^{obs}_n ( R_k) \cos n\varphi_j
+b^{obs}_n ( R_k) \sin n\varphi_j \right] \sin i \right)^2\,,
\end{equation}
where $k$ designates the number of the ring and summing is performed over
all pixels belonging to the ring.

\begin{figure}

\psfig{figure=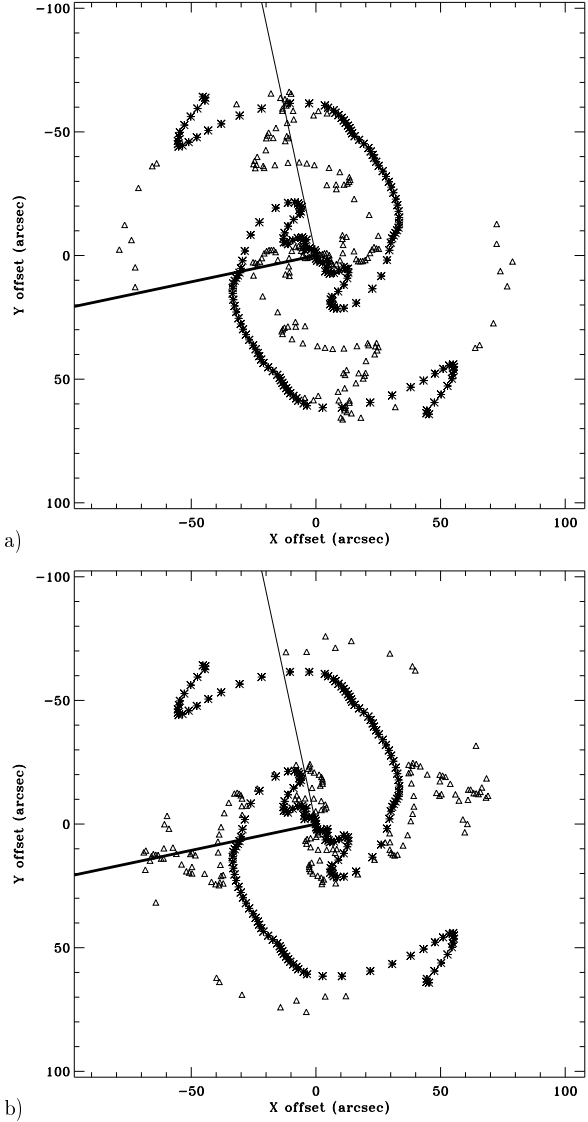,width=8cm,
bbllx=105pt, bblly=95pt, bburx=430pt, bbury=710pt, clip=}

\caption{Comparison of the positions of the maxima (a) and minima (b) of
the ``modified third harmonic'' of the \ha\ line-of-sight velocity field
(triangles) with the form of the spiral arms characterized by positions
of maxima of the second Fourier harmonic of $R$ brightness map
(asterisk). The modified third harmonic is $\cos(2\varphi -
F_3+\pi/2)$, where $F_3$ is the phase of the original third harmonic.}
\label{f-5}
\end{figure}

The first harmonic of the observed velocity field contains contribution
from both the rotation velocity and the perturbed motion, which can not be
separated without taking some additional proposition, but the third
harmonic should be unambiguously related to the velocity
perturbation connected with the observed spiral arms. To verify this,
Fridman et al. (1997) proposed to use the ``modified third harmonic", which
has a form $\cos(2\varphi-F_3+\pi/2)$, where $F_3$ is the phase of the
original third harmonic of the observed line-of-sight velocity. In the
case of a tightly wound spiral, it was shown that the maxima of this
"modified third harmonic" follow the maxima of the perturbed surface
density of the disc outside corotation (that is $F_3$ $=$
$F_\sigma+\pi/2$). Inside corotation, they can coincide with maxima or
minima of the perturbed surface density, depending on the relation between
the amplitudes of the radial and azimuthal residual velocities. If $C_r$
$<$ $C_\varphi$ then $F_3$ $=$ $F_\sigma+\pi/2$, otherwise $F_3$ $=$
$F_\sigma-\pi/2$. Fig.~\ref{f-5} shows the correlation between the
positions of maxima of the perturbed surface density (characterized by the
$m_{\rm obs}~=~2$ harmonic in the $R$-band brightness distribution) and
maxima (a) and minima (b) of the ``modified third harmonic''. As one can
see, the latter set of maxima really follow the spiral
arms in the outer part of the disc and in the very inner region, whereas
they lie between the arms in the radial region $25''$ $<$ $R$ $<$ $40''$.
This proves the correctness of the assumption we used on the wave nature
of the spiral structure in NGC~3631, and implies that the position of
corotation is at about $40''$. This argumentation gives the first rough
estimate of the corotation radius $R_{\rm c}$.

\begin{figure}

\psfig{figure=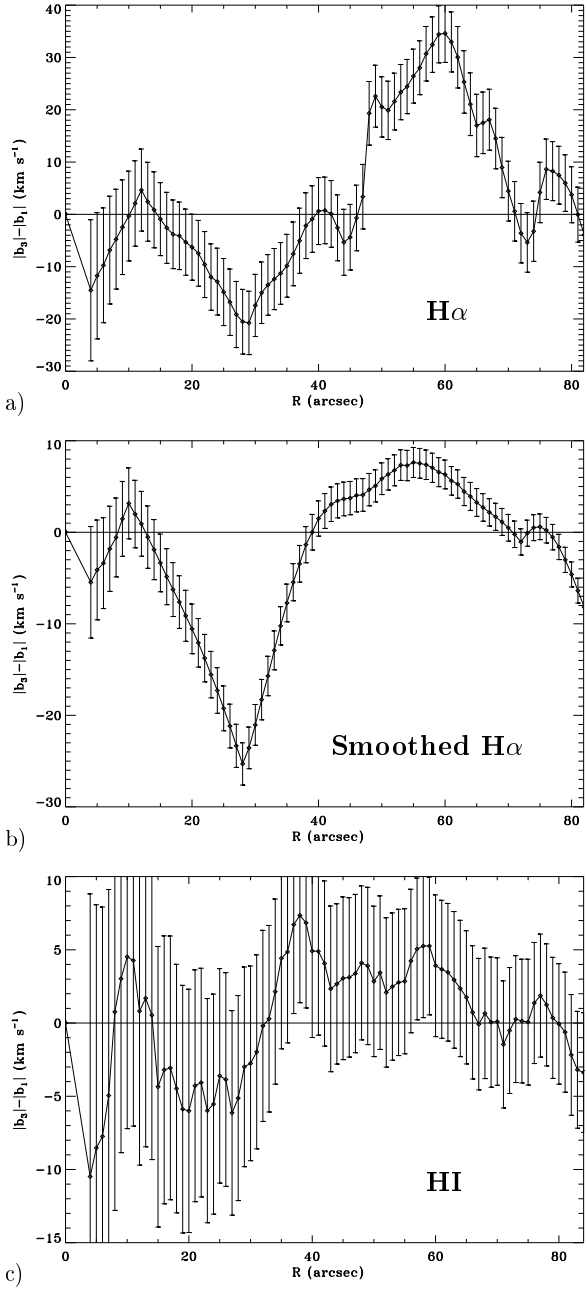,width=8cm,
bbllx=105pt, bblly=45pt, bburx=410pt, bbury=705pt, clip=}

\caption{ Behaviour of $|b_3^{obs}(R)|-|b_1^{obs}(R)|$ as a function of
galactocentric radius $R$ in NGC~3631, as derived from (a) original \ha\
line-of-sight velocity field, (b) smoothed \ha\ velocity field, (c) 21
cm velocity field. Error bars correspond to $3\sigma$ level. According to
results of the density wave theory in the approximation of tightly wound
spirals, the difference should be negative inside corotation and positive
outside. Thus the data show that the corotation radius is at 40$''$ $\pm$
7$''$. }
\label{f-7}
\end{figure}

\begin{figure}

\psfig{figure=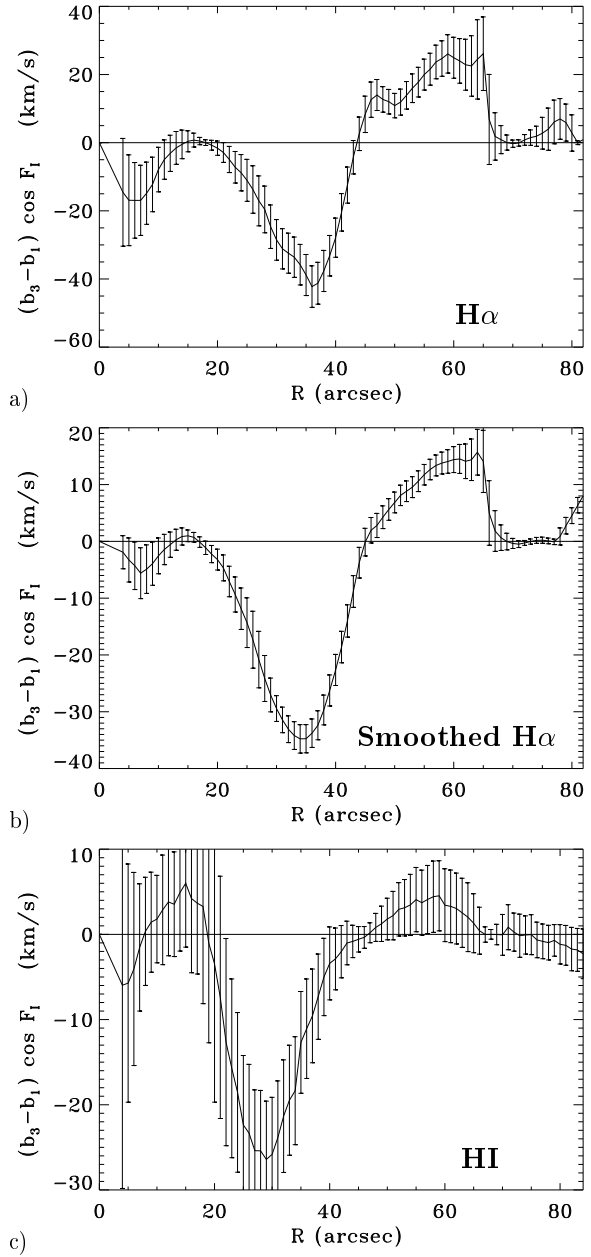,width=8cm,
bbllx=105pt, bblly=85pt, bburx=400pt, bbury=705pt, clip=}

\caption{Variation of $(b_3^{obs}-b_1^{obs}) \cos F_\sigma $ with
galactocentric radius $R$ in NGC~3631, as derived from (a) original \ha\
line-of-sight velocity field, (b) smoothed \ha\ velocity field, (c) 21
cm velocity field. Error bars correspond to $3\sigma$ level. According to
results of the density wave theory in the approximation of tightly wound
spirals, the difference is negative inside and positive outside
corotation. Thus the data show the corotation radius to be about 43$''$
$\pm$ 4$''$.} \label{f-8} \end{figure}

To obtain a more accurate estimate of the position of corotation we
previously proposed two methods (Lyakhovich et al. 1997; Fridman et al.,
1997). The first method is based on the comparison of the radial
behaviour of the sinus components of the first ($b_1^{obs}$) and third
($b_3^{obs}$) Fourier harmonics of the line-of-sight velocity
field. In the case of tightly wound spirals, the following relations
are fulfilled (for details see Lyakhovich et al. 1997 and Fridman et al.,
1997):
\begin{equation}
\label{eq:b1b3}
\begin{array}{ll}
|b_3^{obs}(R)|-|b_1^{obs}(R)| \le 0, & \hbox{for~~ } R < R_c \,, {\rm and} \\
|b_3^{obs}(R)|-|b_1^{obs}(R)| \ge 0, & \hbox{for~~ } R > R_c \, .
\end{array}
\end{equation}

These inequalities enable the determination of the location of the
corotation radius from observational data. According to
Eq.(\ref{eq:b1b3}), corotation is located in the region where the
difference $|b_3^{obs}(R)| - |b_1^{obs}(R)|$ changes sign, from minus to
plus.

Fig.~\ref{f-7} shows the radial behaviour of $|b_3^{obs}(R)| -
|b_1^{obs}(R)|$ in NGC~3631, as obtained from the original \ha\ (a),
smoothed \ha\ (b), and 21-cm (c) line-of-sight velocity
fields. Fig.~\ref{f-7} shows that this function derived from \ha\
line-of-sight velocity field is negative within the errors in the inner
part of the galaxy, and positive in the outer region, in accordance with
the expectations from (\ref{eq:b1b3}). From these data it follows that the
corotation radius is at $R=40''\pm7''$. The level of errors in 21-cm data
is relatively higher, due to lower resolution.
Nevertheless, behaviour of $|b_3^{obs}(R)| - |b_1^{obs}(R)|$ derived from
21-cm line-of-sight velocity is in good agreement with \ha\ data within
one $\sigma$ level of confidence.

In addition, yet another method to estimate $R_{\rm c}$ was proposed
(Lyakhovich et al. 1997; Fridman et al. 1997), based on the relation
between the phases of the perturbed surface density and the radial
perturbed velocity of a gas, which is fulfilled for trailing tightly
wound spirals:

\begin{equation}
F_\sigma-F_r~= \left\{ \begin{array}{ll}
\!\! \pi & \!\!{\rm at} ~R<R_c \\
\!\! 0 & \!\!{\rm at} ~R>R_c
\end{array}
\!\! \right. = ~(1-{\rm sign}\hat \omega) \,\pi/2 \, .
\label{vst8}
\end{equation}

This relation shows that gas should move along spiral arms inwards
inside of corotation, and outwards outside of it. As shown in the cited
papers, it may also be written as the relationship between the phase of
the perturbed surface density and the Fourier coefficients $b_3^{obs}$
and $b_1^{obs}$:

\begin{equation}
\label{eq:new2}
\begin{array}{ll}
(b_3^{obs}(R) - b_1^{obs}(R))\, \cos F_\sigma (R) ~\le~ 0, &
~~ \hbox{for~} R < R_c, \\
(b_3^{obs}(R)-b_1^{obs}(R)) \, \cos F_\sigma (R) ~\ge~ 0, &
~~ \hbox{for~} R > R_c.
\end{array}
\end{equation}

Fig.~\ref{f-8} shows the dependence of $(b_3^{obs}-b_1^{obs}) \cos
F_\sigma $ on galactocentric radius $R$. The situation is quite similar
to presented in the previous figure. The current approach leads to the
estimate of $R_{\rm c}$ of $43''\pm4''$.

Combining the results from these methods, which agree
well, we may conclude that the corotation radius $R_c \approx 42'' \pm
5''$, or 3.2~kpc$\pm$0.38~kpc.

\section{Restored velocity field of gas in the galactic plane}

As shown in Section~2, lowering the angular resolution leads to an
underestimate of the amplitudes of the Fourier harmonics. Therefore, to
restore a two component vector velocity field of a gas in the disc of
NGC~3631, we use the full resolution \ha\ line-of-sight velocity field.

To restore the velocity field in the plane of a galactic disc, i.e., its
radial $\tilde V_r $ and azimuthal $V_\varphi = V_{rot} + \tilde
V_\varphi$ velocity components, it is necessary to determine five
unknown functions: $V_{rot}(r)$, $C_r(r)$, $C_\varphi (r)$, $F_r(r)$,
$F_\varphi (r)$ (see Eqs.~\ref{eq:sprvz1} and \ref{eq:sprvz2}). These
five functions are connected with the Fourier coefficients of the
observed line-of-sight velocity field by the four relations
(\ref{eq:aAbB1}), (\ref{eq:aAbB2}), (\ref{eq:aAbB5}), and
(\ref{eq:aAbB6}). An additional condition, required to close the system,
should have a theoretical origin. Unfortunately, up to now, a reliable
condition, valid for any density wave amplitude, is not available.
Several possibilities discussed in the literature (Sakhibov \& Smirnov,
1989; Fridman et al. 1997) are based on  some kind of
approximation and have limited applicability. To overcome this
difficulty, we propose the following approach.

Among the functions listed above, $V_{rot}(r)$ can be most reliably
estimated in two independent ways from observational data. The first one
uses the equilibrium condition of a gaseous disc rotating in a
gravitational potential $\Psi$

\begin{equation}
V^2_{rot}/r ~=~ {\partial \Psi}/{\partial r} \,.
\label{19}
\end{equation}

The right-hand side of the Eq.(\ref{19}) is determined from the mass
distribution in a galaxy, or its surface brightness maps, assuming the
mass-to-light ratio is known and constant with radius. For this purpose,
we use a three-component dynamical model of a spiral galaxy similar to the
one used by Sumin, Fridman, \& Haud (1991). Although the model is rather
crude, the resulting rotation curve $V_{rot}(R)$ corresponds to the region
approximately between $(a_1)_{\min}$ and $(a_1)_{\max}$.

\begin{figure}
\vspace{-.2cm}
\psfig{figure=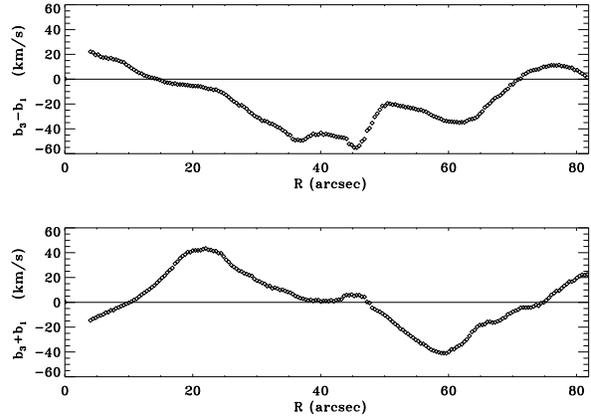,width=8.3cm}

\vspace{-.2cm}

\caption{The radial dependence of $b_3^{obs}-b_1^{obs}$ $=$ $C_r
\, \cos \,F_r$ and $b_3^{obs}+b_1^{obs}$ $=$ $C_\varphi \, \sin
\,F_\varphi$ observed in the spiral galaxy NGC~3631. An estimate of
the amplitudes of the velocity components from the extremes of these
functions gives $max(C_r)$ $\simeq$ $max(C_\varphi)$ $\simeq$ 60~km/s.}
\label{f-v-2}
\end{figure}

The same result can be obtained in another way. From equation
(\ref{eq:aAbB1}) it follows that the difference $|a_1-V_{rot}|$ cannot
exceed the amplitudes $C_r$ and $C_\varphi$, which, in turn, are
connected by equations (\ref{eq:aAbB2}), (\ref{eq:aAbB5}), and
(\ref{eq:aAbB6}) with the Fourier coefficients $b_1^{obs}$, $a_3^{obs}$,
and $b_3^{obs}$, determined from our observations. In Fig.~\ref{f-v-2},
we show the radial behaviour of $b_3^{obs}-b_1^{obs}$ $=$ $C_r \, \cos
\,F_r$ and $b_3^{obs}+b_1^{obs}$ $=$ $C_\varphi \, \sin
\,F_\varphi$. The extremes of these functions allow estimates of the
amplitudes $C_r$ and $C_\varphi$. From Fig.~\ref{f-v-2} we conclude
that in NGC~3631 a maximum value of the amplitude of the residual
velocities occurs at $60$~km/s, i.e.

\begin{equation}
|a_1 - V_{rot}|_{\max} \le 60 {\rm ~km/s}.
\label{20}
\end{equation}

\begin{figure}
\hspace{.2cm}\psfig{figure=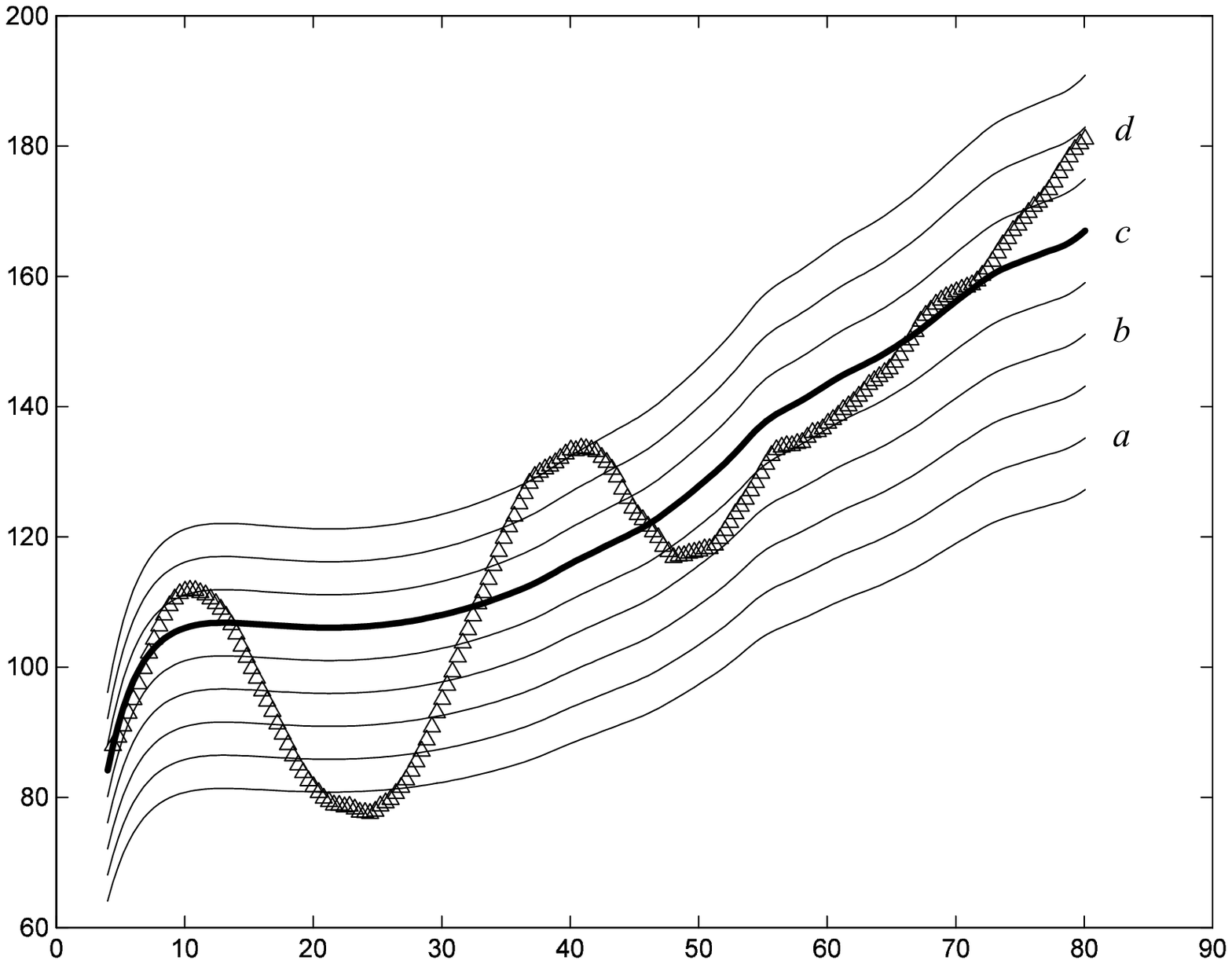,width=8.0cm,
bbllx=72pt, bblly=210pt, bburx=548pt, bbury=585pt, clip=}

\vspace{-.1cm}

\caption{Examples of trial curves used to represent the rotation curve
($V_{rot}(r)$) in NGC~3631, shown by solid lines, along with the observed
behaviour of $a_1^{obs}(r)$ (triangles). Abscissa is the radius in arcsec
and ordinate --- rotation velocity in km/s. According to the analysis
presented below, the thickest line marks the curve corresponding to the
real rotation curve of NGC~3631 gaseous disk.}
\label{f-v-1}
\end{figure}

The conditions (\ref{19}) and (\ref{20}) do not allow an exact
calculation of the function $V_{rot}(r)$. Nevertheless, they set limits
on the variations of both the amplitude and the form of
$V_{rot}(r)$. Within these limits, varying the mass-to-light ratio ($\pm
40\%$) and assuming it does not depending on radius we obtain a set of
trial curves (Fig.~\ref{f-v-1}) and analyse the velocity fields restored
from Eqs.~(\ref{eq:aAbB1})--(\ref{eq:aAbB6}) for a given $V_{rot}(r)$.

\begin{figure*}
\psfig{figure=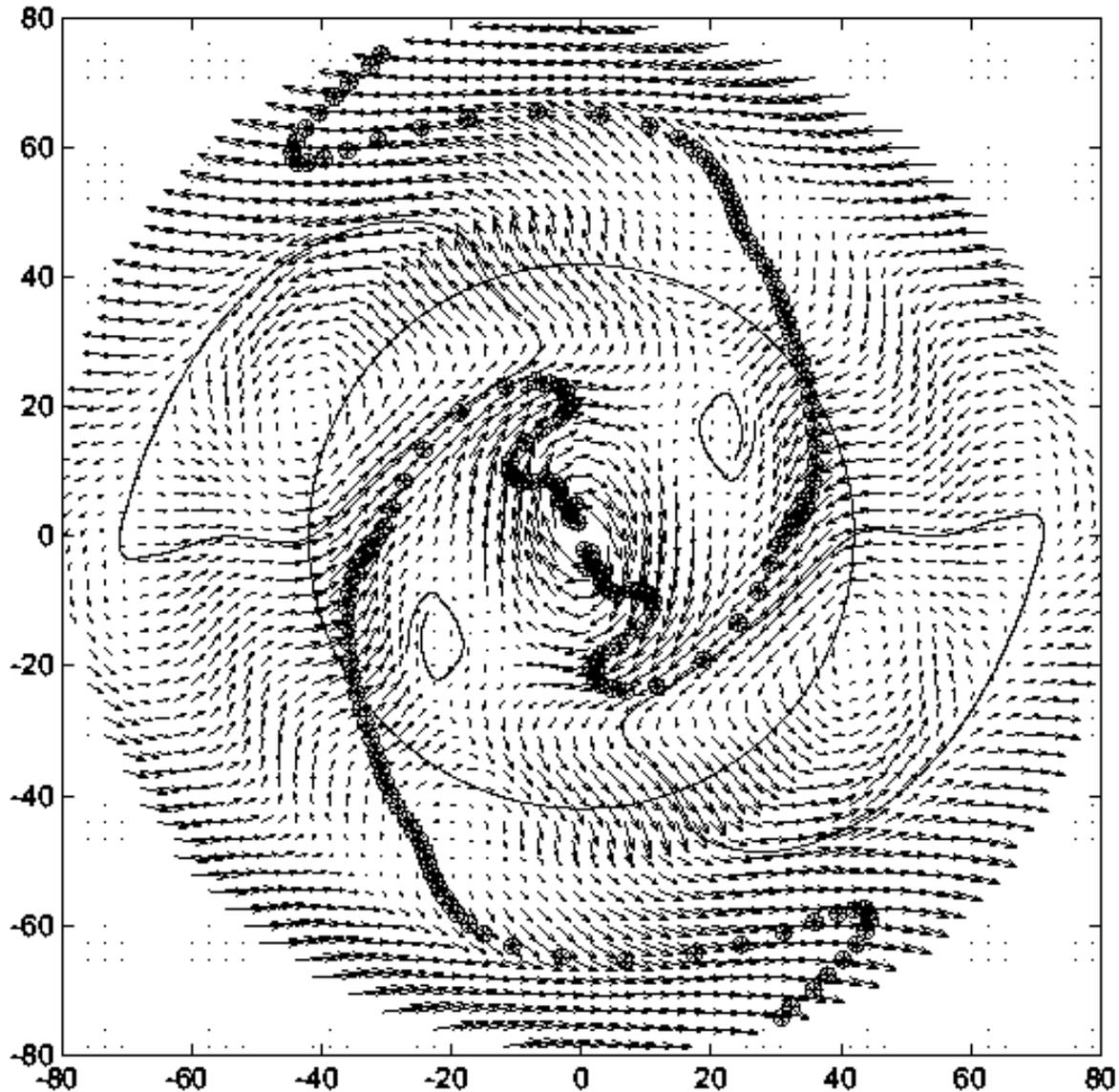,width=17cm,
bbllx=120pt, bblly=210pt, bburx=500pt, bbury=585pt, clip=}

\caption{Restored velocity field of NGC~3631 in the plane of the disc
in the reference frame rotating with the pattern speed. Overlaid
asterisks show the locations of maxima of the second Fourier harmonic of
the $R$-band brightness map of the galaxy. The circle marks the
position of the corotation. Solid lines demonstrate vortex separatrices or
nearly closed streamlines in the absence of a separatrix (the ones greater
in size correspond to anticyclones, and the smaller -- to cyclones).
\newline (a) The curve marked by symbol "a" in Fig.~11 used as
$V_{rot}(r)$.} \label{f-9aa} \end{figure*}

\setcounter{figure}{11}

\begin{figure*}
\psfig{figure=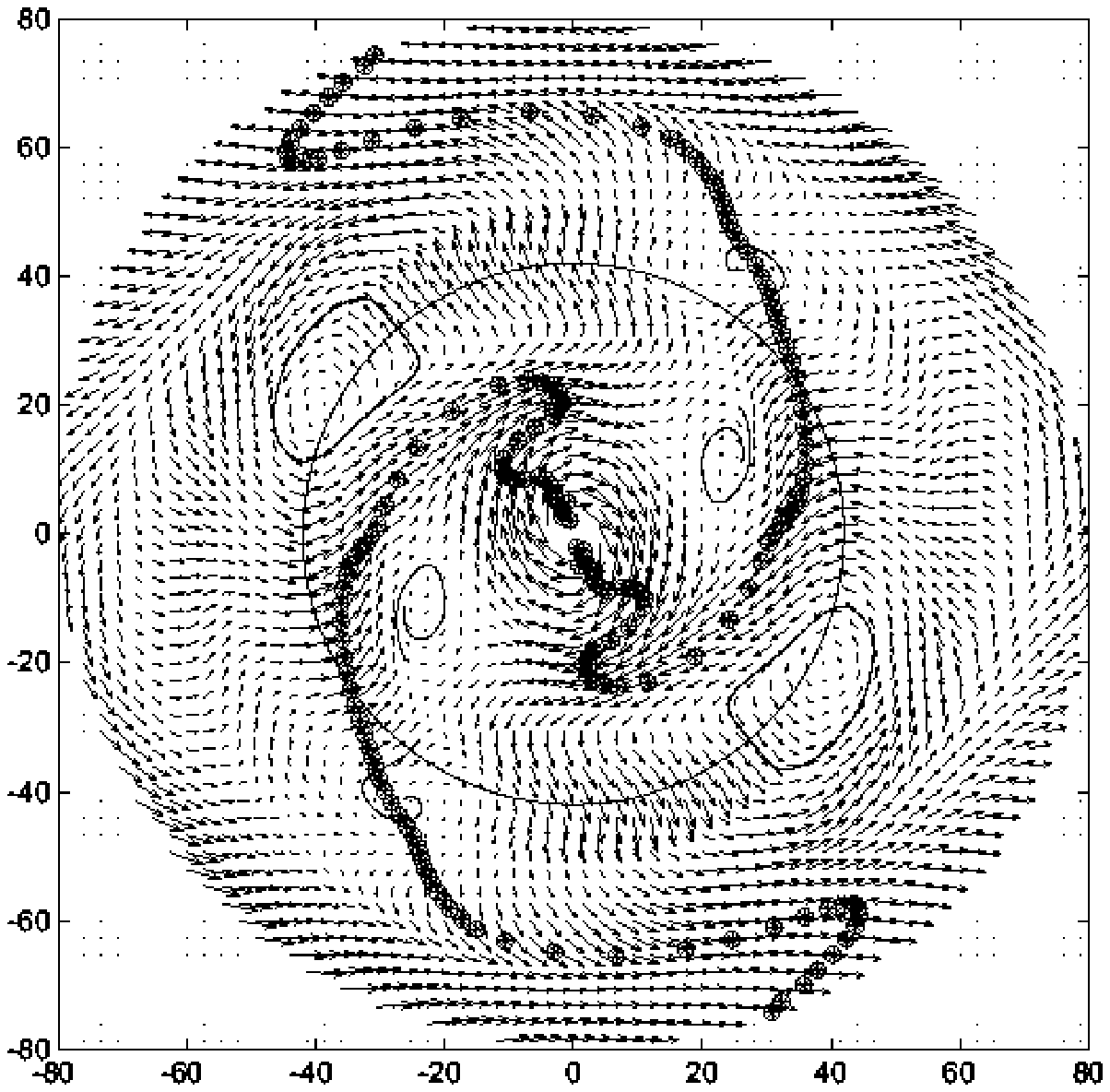,width=17cm,
bbllx=120pt, bblly=210pt, bburx=500pt, bbury=585pt, clip=}

\caption{(b) The curve marked by symbol "b" in Fig.~11 is used as
$V_{rot}(r)$.}
\end{figure*}

\setcounter{figure}{11}

\begin{figure*}
\psfig{figure=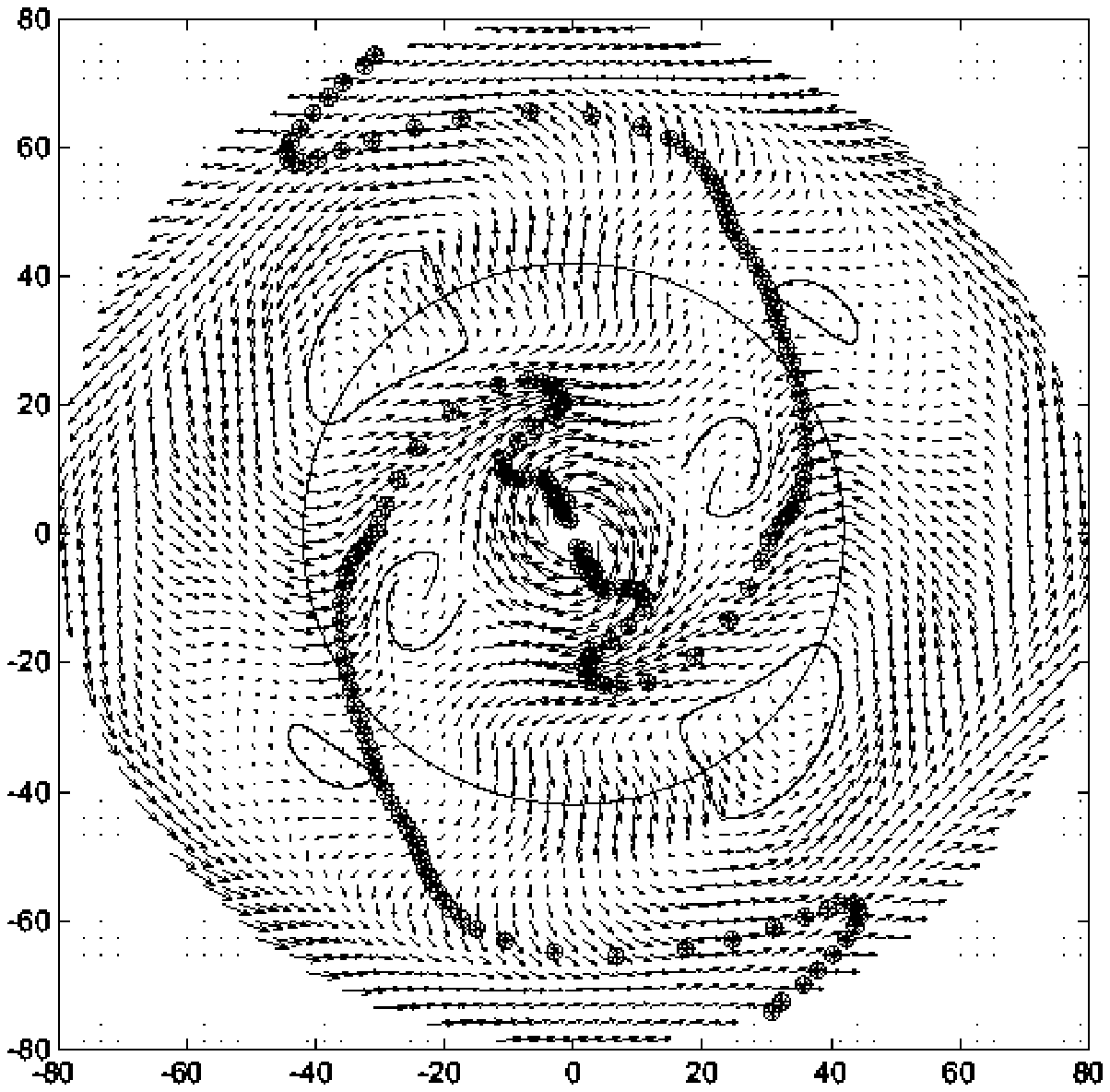,width=17cm,
bbllx=120pt, bblly=210pt, bburx=500pt, bbury=585pt, clip=}

\caption{(c) The curve marked by symbol "c" in Fig.~11 is used as
$V_{rot}(r)$.}
\end{figure*}

\setcounter{figure}{11}

\begin{figure*}
\psfig{figure=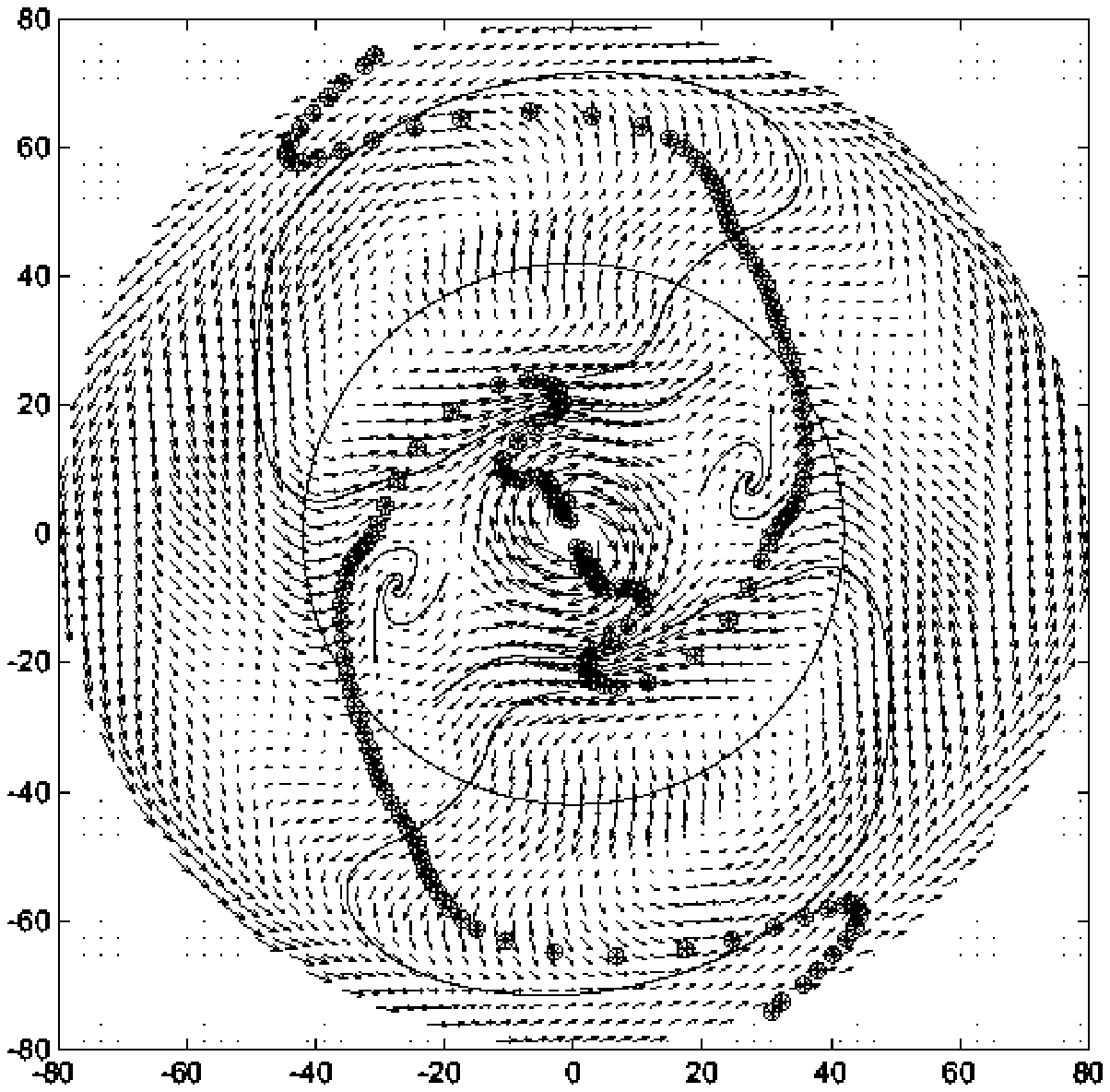,width=17cm,
bbllx=120pt, bblly=210pt, bburx=500pt, bbury=585pt, clip=}

\caption{(d) The curve marked by symbol "d" in Fig.~11 is used as
$V_{rot}(r)$.} \end{figure*}

Fig.~\ref{f-9aa} shows four examples of the restored velocity field of
NGC~3631, in the plane of the disc, and in a reference frame rotating
with the spiral pattern. Positions of the maxima of the second Fourier
harmonic of the $R$-band surface brightness map are overlaid to indicate
the relative location of spiral arms and structures in the velocity
field.

It is clearly seen that, in spite of broad variations in $V_{rot}(r)$,
the general structure of the velocity field changes only slightly. In
all cases this structure demonstrates the presence of two anticyclones,
located on the corotation circle, and between spiral arms. Variations
of the rotation curve only lead to small changes in the basic
quantitative features of the anticyclones. In all cases, the maximum
noncircular velocity in the vortices is about 60~km/s, their radial
width is about 20$''$, and the variation of the azimuthal position of
the centres of anticyclones is less than 10$^\circ$. This proves that
the exact shape of the rotation curve does not impact on the basic
conclusion of the existence of giant anticyclones in the gaseous disc of
NGC~3631. These anticyclones are similar to those revealed earlier in
the velocity field of NGC~157 (Fridman et al. 1997).

Another new result is that regions of cyclonic shear are seen in
Fig.~\ref{f-9aa}. As a consequence of the relatively high amplitude of
the density wave in NGC~3631, a cyclonic shear is produced by the
density wave in some regions, which dominates the anticyclonic shear
caused by differential rotation. The appearance of cyclones in gaseous
galactic discs with a strong density wave was predicted
earlier by Fridman et al. (2000).

To choose between the velocity fields presented in Fig.~\ref{f-9aa} the
field which is closest to the real velocity field of the galactic disk
of NGC~3631, we use the following criterion. In the course of many
revolutions the growth of the density wave amplitude is stopped by the
saturation of a corresponding instability. Thus in the reference frame
corotating with spiral arms the velocity field of the galactic disk should
be stationary. In such a field the vortices should have closed
separatrices (lines dividing two families of trajectories:  trapped and
transit ones). According to the criterion stated above, the velocity field
presented in Fig.~\ref{f-9aa}\,c is distinct in that sense that it is only
case where cyclones have close separatrices. It is an argument in favour
of choosing just this example as the closest to the real velocity field
of the NGC~3631 gaseous galactic disk. At the same time that means that the
curve "c" in Fig.~\ref{f-v-1} is close to the real rotation curve of the
disk. The relative position of the vortices and spiral arms in
Fig.~\ref{f-9aa}\,c is in a good agreement with theoretical predictions
(Fridman et al. 1999).

\section {Conclusions}

We can briefly summarize our main conclusions from this paper as
follows:

\begin{enumerate}

\item Based on an analysis of line-of-sight velocity fields of gaseous
emission lines for the spiral galaxy NGC~3631 we confirm observationally
the theoretical conclusions from our previous work about the wave nature
of its two-armed spiral structure, and discuss its origins and
properties in light of this theoretical framework.

\item Using two independent methods, we find that the
corotation radius in this galaxy is at about 42$''$ or 3.2~kpc.

\item The projection of the restored three-dimensional vector velocity
field of gas in the plane of the galaxy, and in a reference frame
corotating with the spiral pattern, reveals the presence of two
anticyclonic vortices near corotation. We thus confirm the theoretical
prediction foreseen earlier on the basis of a study of the general
principles of the wave nature of spiral structure in galaxies.

\item We show the existence of cyclonic vortices in NGC~3631, apart from
that of anticyclones, as mentioned above. Such cyclonic vortices are a
consequence of a high amplitude of the density wave in this galaxy.

\end{enumerate}

{\it Acknowledgements} This work was performed with partial financial
support of RFBR grant N 99-02-18432, ``Leading Scientific Schools"
grant N 00-15-96528, and ``Fundamental Space Researches. Astronomy"
grants: N 1.2.3.1, N 1.7.4.3. The Jacobus Kapteyn
Telescope is operated on the island of La Palma by the Isaac Newton
Group in the Spanish Observatorio del Roque de los Muchachos of the
Instituto de Astrof\'\i sica de Canarias. Data were retrieved from the ING
archive.

\end{document}